\documentclass[twocolumn,showpacs,preprintnumbers,amsmath,amssymb]{revtex4}
\usepackage{graphicx}% Include figure files
\usepackage{dcolumn}% Align table columns on decimal point
\usepackage{bm}% bold math

\newcommand{\ket}[1]{|#1\rangle}
\newcommand{\bra}[1]{\langle#1|}
\newcommand{\abs}[1]{|#1|}

\renewcommand{\i}{{\rm i}}
\renewcommand{\d}{{\rm d}}
\newcommand{\e}{{\rm e}}

\newcommand{\tS}{\widetilde{s}}

\begin{document}

\preprint{QIP07WW1\_deep\_donors}

%\preprint{APS/123-QED}

\title{Exchange between deep donors in semiconductors: a quantum defect approach}
\author{W.~Wu}
\email{wei.wu@ucl.ac.uk}
\author{A.J.~Fisher}
 \email{andrew.fisher@ucl.ac.uk}
\affiliation{UCL Department of Physics and Astronomy and London Centre for Nanotechnology,\\
University College London, Gower Street, London WC1E 6BT}
\date{\today}%
\begin{abstract}
Exchange interactions among defects in semiconductors are commonly
treated within effective-mass theory using a scaled hydrogenic
wave-function.  However such a wave-function is only applicable to
shallow impurities; here we present a simple but robust
generalization to treat deep donors, in which we treat the
long-range part of the wavefunction using the well established
quantum defect theory, and include a model central-cell correction
to fix the bound-state eigenvalue at the experimentally observed
value.  This allows us to compute the effect of binding energy on
exchange interactions as a function of donor distance; this is a significant quantity given recent
proposals to carry out quantum information processing using deep
donors.  As expected, exchange interactions are suppressed (or
increased), compared to the hydrogenic case, by the greater
localization (or delocalization) of the wavefunctions of deep
donors (or `super-shallow' donors with binding energy less
then the hydrogenic value). The calculated results are compared
with a simple scaling of the Heitler-London hydrogenic exchange;
the scaled hydrogenic results give the correct order of magnitude
but fail to reproduce quantitatively our calculations. We
calculate the donor exchange in silicon including inter-valley
interference terms for donor pairs along the $\{100\}$ direction,
and also show the influence of the donor type on the distribution
of nearest-neighbour exchange constants at different
concentrations. Our methods can be used to compute the exchange
interactions between two donor electrons with arbitrary binding
energy.
\end{abstract}

\maketitle
\section{Introduction}\label{sec:whittaker}
Accurate estimations of exchange interactions among semiconductor
defects are crucial in determining the magnetic properties of
doped semiconductors near to a metal-insulator transition
\cite{andres} and in assessing the potential of such defects for
potential applications in quantum information processing
\cite{kane,sfg}.  However, the long-range tails of defect
wavefunctions make it difficult or impossible to treat these
systems using fully \textit{ab initio} techniques, so such
calculations are generally performed within empirical models such
as the well-established effective mass approximation
\cite{luttinger,kittel,kl55}.  The quantum-mechanical problem for
a single defect is thereby reduced to that of a single electron
moving in an effective medium determined by the effective mass and
static permittivity of the host crystal; in the simplest form of
the theory, the solution becomes that of a scaled hydrogen atom.
The exchange between two such defects can then be obtained from
well-established treatments of exchange in the hydrogen molecule,
ranging from the simple Heitler-London model \cite{heitlerlondon}
to the more sophisticated approach of Herring and Flicker
\cite{herring62,herringflicker} which exactly accounts for the
two-electron correlations in the limit of large separations.

However, a number of complications arise when considering real
defect systems. First, the minima of the conduction band may not
be isotropic; this is so in the important case of silicon, where
there are six degenerate minima lying along the $\{100\}$
directions in the Brillouin zone and each having a significant
anisotropy ($m_\perp=0.98 m_e$, $m_\parallel=0.19 m_e$), where $m_e$ is
electron mass. This is commonly dealt with either by simply
adopting an effective isotropic dispersion with an appropriate
average effective mass (for example, $m^*=(m_\parallel
m_\perp^2)^{1/3}$), or by making a variational \textit{ansatz} for
the wavefunction of the anisotropic system, such as that of
Luttinger and Kohn \cite{luttinger,kittel,kl55}.

Second, the existence of more than one minimum needs to be
accounted for; this results in an effective-mass wavefunction of
the form
\begin{equation}\label{eq:multibandeffmass}
\psi(\vec{r})=\sum_n \alpha_nF_n(\vec{r})\phi_{n\vec{k}_0}(\vec{r}),
\end{equation}
where the sum runs over the different minima, $F_n$ is a
slowly-varying envelope function, and
$\phi_{n\vec{k}_0}(\vec{r})=\exp(\i\vec{k}_0\cdot\vec{r})u_{n\vec{k}_0}(\vec{r})$
is the Bloch function for the $n$th minimum. The coefficients
$\alpha_n$ arise from the coupling of the different band minima by
short-range (large-wavevector) components of the potential; for an
isolated substitutional donor in a perfect crystal, they
correspond to an irreducible representation of the $T_d$ point
group. In this paper we concentrate on the ground state of defects
in silicon, for which the terms in
equation~(\ref{eq:multibandeffmass}) correspond to the six
$\{100\}$ directions and the appropriate representation is the
identity representation:
\begin{eqnarray}\label{eqn:a0toa5}
\alpha^{(0)}&=&\frac{1}{\sqrt{6}}(1,1,1,1,1,1).
%\\\alpha_1&=&\frac{1}{\sqrt{12}}(1,1,1,1,-2,-2)
%\\\alpha_2&=&\frac{1}{2}(1,1,-1,-1,0,0)
%\\\alpha_3&=&\frac{1}{\sqrt{2}}(1,-1,0,0,0,0)
%\\\alpha_4&=&\frac{1}{\sqrt{2}}(0,0,1,-1,0,0)
%\\\alpha_5&=&\frac{1}{\sqrt{2}}(0,0,0,0,1,-1)
\end{eqnarray}

The existence of these distinct parts to the wavefunction gives
rise to interference terms in properties of defect pairs as a
function of their separation. Such terms were originally
considered in electron hopping and recombination rates \cite{miller60,enck69}, and later
included in exchange interactions \cite{cullis70,andres}. Other
more recent calculations are discussed below. Note, however, that
the expressions used in \cite{andres} to average the exchange over
the different minima assume a particular physical origin for the
exchange (namely the direct electron-electron Coulomb interaction
term) which does not, in fact, dominate the problem.  Although its
magnitude has approximately the same asymptotic scaling as the
full exchange, by itself the direct term would lead to a
ferromagnetic spin-spin interaction.  The observed interaction is
always antiferromagnetic, as it must be for a two-electron system
\cite{heisenberg}).

Both the anisotropy and multiple-minima problems can be solved
while remaining largely or entirely within the framework of
effective-mass theory, and a number of recent papers have
addressed the exchange between hydrogenic donors by this method.
Much of the inspiration for this work came from the proposal
\cite{kane} to use such exchange in shallow donors as a
qubit-qubit coupling mechanism to drive entangling gates in
quantum information processing. The importance of precise donor
positioning in determining the value of the exchange was
emphasized in \cite{koiller01}, but this paper follows
\cite{andres} in using only the (ferromagnetic) direct exchange
interaction for the calculations. The full Heitler-London formula
for the exchange was used in a following paper \cite{koiller02},
where the contribution of strain to modifying the interaction was
also discussed.  These calculations were performed by neglecting
the spatial variation of the periodic part of the Bloch functions
$u_{n,\vec{k}_0}$ appearing in
equation~(\ref{eq:multibandeffmass}); this approximation was
carefully examined in \cite{wellard03} and found to be very
accurate, and this finding was confirmed by a subsequent
calculation of the electronic structure of defect pairs in which
the Bloch functions from an \textsl{ab initio} treatment of the
host crystal were used \cite{koiller04}.  The same paper also
found that allowing the phases of the different Bloch function
contributions to `float' relative to one another in the case of
the defect pair made small difference to the results.

The third difficulty is in many ways most serious: effective-mass
theory predicts that all single-electron donors in a given host
should have the same binding energy, independent of their chemical
nature.  This is far from the case: in silicon, for example,
observed binding energies range from 31.2\,meV for Li and
45.6\,meV for P through to 71.0\,meV for Bi \cite{ramdas81},
whereas the calculated value using the Luttinger-Kohn variant of
effective mass theory is 31.3\,meV \cite{ramdas81}. It is very
desirable to predict exchange in deeper donors by similarly simple
methods: they are advantageous for quantum information processing
\cite{sfg} because they are less prone to ionization, and have
much longer spin-lattice relaxation lifetimes
\cite{castner63}---indeed, resonant Raman routes for spin
relaxation are completely absent if the level spacing exceeds the
maximum phonon energy of the host material (64.5\,meV for Si).

Treating deep donors requires some significant corrections to
effective mass theory. Two common approaches involve making an
explicit short-range correction to the Coulomb field of the
impurity (the so-called `central cell correction'), or altering
the long-range solution in the Coulomb field so that it
corresponds to the observed binding energy (the `quantum defect'
approach, which has its origins in atomic physics)
\cite{bebb,seaton}.  However, the effect of these modifications on
the exchange interactions of the defects has remained largely
unknown.

In this paper we show how the exchange in the most physically relevant range of separations can be computed between
donors with arbitrary binding energies (including the important case of deep donors). We
use a quantum-defect description, with a simple model potential to
represent the central-cell correction; for reasons that are
explained below, it is important to compute the exchange using a
potential for which the wavefunction concerned is an
eigenfunction.  We first introduce the techniques that we use: the
effective-mass model, quantum-defect corrections to it and the
simple central-cell corrections, followed by the methods we use to
calculate exchange. Then we present our results for defects of
different binding energies.

\section{Method of calculation}
\subsection{Effective-mass theory}
The effective-mass equation \cite{luttinger,kittel,kl55} reads:
\begin{eqnarray}
[\epsilon_n(\vec{k}_0+\frac{1}{i}\nabla)+U]F_n=\epsilon F_n.
\end{eqnarray}
where it is intended that the band energy $\epsilon_n$ be expanded
around the band extremum $\vec{k}_0$ to second order-terms in $(1/i)\nabla$.  $F_n$
is the envelope function, in terms of which the true wavefunction
is expanded using
\begin{eqnarray}\label{eq:effectivemasssum}
\psi=\sum_n \alpha_n F_n(\vec{r})\phi_{n\vec{k}_0}(\vec{r}),
\end{eqnarray} where $F_n$ is a solution of the effective mass
equation.

In the simplest theory, the effective mass tensor is replaced by a
single averaged effective mass $m^{*}$, resulting in an effective
isotropic equation for the envelope function, which is then independent of the index $n$:
\begin{equation}\label{eq:effmasseqn}
[-\frac{\hbar^2}{2m^{*}}\nabla^2-\frac{e^2}{\epsilon_r
r}-\epsilon]F(\vec{r})=0,
\end{equation}
where $\epsilon_r$ is the relative permittivity of the host.
In this paper we will follow other recent treatments (for example
\cite{koiller01,koiller02}) and work with this isotropic equation
as our starting point.  For silicon, $m^*=0.33 m_e$ and $\epsilon_r=11.7$; this leads to a
set of scaled atomic units for the hydrogenic impurity problem (length $a_0^*=1.94\mathrm{nm}$,
energy $\mathrm{Ha}^*=0.062\mathrm{eV}$).

\subsection{The quantum defect method}
The first component of our correction to effective-mass theory
involves treating the wavefunction far from the impurity by the
quantum defect method \cite{seaton}.  Even in the simplest isotropic
approximation, the method was found to provide quantitative
results giving good approximations to both the observed spectral
dependence and magnitude of the photo-ionization cross section \cite{bebb}. An
essential feature of the quantum defect method is that good
approximate wave functions valid in the region outside the
impurity ion core can be determined using only a knowledge of the
energy eigenvalues. Rather than attempting to solve it
 as an eigenvalue equation to
determine the allowed spectrum of $\epsilon$, equation~(\ref{eq:effmasseqn}) is
considered valid for large $r$ only and solved for the asymptotic
form of the envelope functions $F(\vec{r})$ corresponding to the
\emph{empirical} value of $\epsilon$. Therefore the method
can deduce the long-range part of the donor wave functions associated with the observed
energy levels,whether deep or shallow, provided that the dominant corrections to 
effective-mass theory are short-range (operate only near the defect).  Of the several different terms believed to contribute to the shift in binding energy for deep donors (see, for example, \cite{stoneham75}), only the electron-phonon interaction operates far from the defect and its contribution is believed to be small.

The most useful form of the far-field solution is a multiple of
the well-known Whittaker function which is just a particular
linear combination of two standard linearly independent confluent
hypergeometric functions, the combination being determined by the
boundary conditions at infinity. We consider the auxiliary radial
function defined by $P(r)\equiv r R(r)$, and write
\begin{equation}\label{eq:whittaker}
P_{\nu,l}(r)=N_{\nu,l}W_{\nu,l+\frac{1}{2}}(2r/\nu),
\end{equation}
where $W$ is a Whittaker function \cite{abramowitzstegun}, $N$ is a normalization
constant, and $\nu$ is the quantum defect parameter defined by
\begin{equation}\label{eq:donorenergy}
\epsilon={- 1\over 2\nu^2}.
\end{equation}
If and only if the donor is hydrogenic (i.e. shallow), $\nu$ is an
integer.

\subsection{Model central cell corrections}\label{sec:centralcell}
For non-integer values of $\nu$, the radial function associated
with the solution~(\ref{eq:whittaker}) diverges like $R(r)=P(r)/r
\sim \frac{1}{r}$ as $r\rightarrow 0$. (Nevertheless this
divergence is integrable, so the functions have a well-defined
normalization for all $\nu$ \cite{seaton}.) In order to find a
solution to equation~(\ref{eq:effmasseqn}) that remains finite at
the origin for a general given energy $\epsilon$, we have to
correct the potential at short distances. One way to do this is to
look for corrections based on the local physics of the impurity
(for example incorporating correctly the transition from a
screened to an unscreened nuclear potential
\cite{nara66,nara67,kerridge}, or including self-consistently the scattering effects of the impurity by means of a pseudoptential based on the microscopic physics \cite{pantelides74}); another way is to correct the
potential at small $r$ empirically solely in order to make the
solution regular there at the experimentally observed energy
eigenvalue.  In this second case the potential will \textit{not} correspond to the physics operating in the core region of the real defect, but it will produce the correct shift in binding energy.  We adopt the second approach in this paper, and refer
to the empirical correction as a model central-cell correction. We
should expect that the two approaches would give similar results
for the calculation of exchange interactions, since they are
determined predominantly by the long-range behaviour of the
wavefunction.

We can then write the Hamiltonian of a single defect center $A$ as
\begin{equation}
\hat{H_{A}}=-\frac{1}{2}\nabla^{2}-\frac{1}{|\vec{r}-\vec{R}_{A}|}+\delta
V(\vec{r}-\vec{R}_{A}),
\end{equation}
where the third term on the right hand side is model central cell
correction, which is not unique. In this work we used two forms
for $\delta V$: a $\delta$-function shell or a square-well
potential.
\subsubsection{$\delta$ potential
correction}\label{subsubsec:delta-ccc}
 Let us choose $\delta V$ to
be a potential "shell" at radius $r=a$, where $a$ is very small
($a \ll 1$ in scaled atomic units):
  \begin{equation}
  V(r)=-\frac{1}{r}+ \lambda\delta(r-a).
  \end{equation}
In the following discussion, we will only consider s-symmetry wave
functions because they are the donor ground states. After matching
the large-$r$ solution to a solution valid for small $r$ that is
regular at the origin and obeys the cusp condition for a nucleus
of charge $Z=1$ ($R'(0)=-R(0)$), we find the solution is
\begin{equation}
  R(r)=\left\{
  \begin{array}{r@{\quad:\quad}l}
  R(a)\frac{1-r}{1-a}&r\leq a\\
  P_{\nu,0}(r)/r&r>a
  \end{array}\right\},
\end{equation}
where $P_{\nu,0}(r)$ is the long-range solution for s-states
defined by equation~(\ref{eq:whittaker}). Figure
(\ref{pic:whitkccsq}) shows the wave function for deep and shallow
donors, both with the central cell correction and without it. Note
how the diverging tails of the Whittaker function are cut off for
$r<a$.
\subsubsection{Square-well potential correction}\label{subsubsec:sw-ccc}
\begin{figure}[htbp]
\begin{tabular}{c}
\includegraphics[width=7cm,height=6cm]{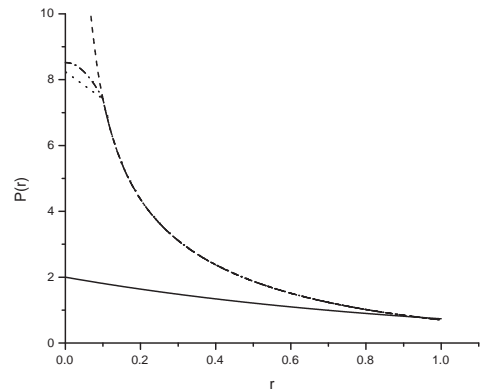}\\
(a)\\
\includegraphics[width=7cm,height=6cm]{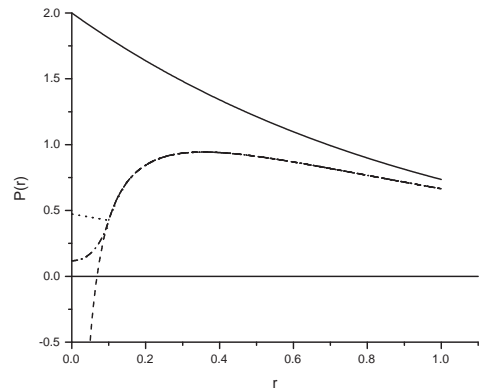}\\
(b)\\
\end{tabular}
 \caption{Comparison of the wavefunctions generated by the
$\delta$ potential  and square-well potential central-cell
corrections. (a) Deep donor, $\nu=0.7$; (b) super-shallow donor,
$\nu=1.1$. In each case, the dotted curve shows $P_{\nu,0}^{\delta}(r)/r$
with a $\delta$-potential central cell correction applied at $a=0.1$, the
dashed-dotted curve is $P_{\nu,0}^{sw}(r)/r$ with a square-well
potential central cell correction applied with $a=0.1$; dashed curve:
uncorrected Whittaker function $P_{\nu,0}/r$ plotted for
comparison; solid curve: radial part of a true $1s$ Hydrogenic
state ($\nu=1$). Notice how the diverging tail of the Whittaker
function is cut off for $r<a$ by both central cell corrections;
the form of the curves for $r<a$ depends on the details of the
potential applied.}\label{pic:whitkccsq}
\end{figure}

An alternative matching is to replace the Coulomb potential by a
square well, of depth $V_{0}$, for $r<a$:
\begin{equation}
  V(\vec{r})=\left\{
  \begin{array}{r@{\quad:\quad}l}
  V_{0}&r\leq a\\
  -\frac{1}{r}&r>a
  \end{array}\right\}
\end{equation}
This model correction gives us:
\begin{itemize}
\item $E>V_0$
\begin{equation}
P(r)=\left\{
  \begin{array}{r@{\quad:\quad}l}
  P_{\nu,0}(a)\frac{\sin{kr}}{\sin{ka}}&r\leq a\\
  P_{\nu,0}(r)&r>a
  \end{array}\right\}
\end{equation}
\item $E<V_0$
\begin{equation}
P(r)=\left\{
  \begin{array}{r@{\quad:\quad}l}
  P_{\nu}(a)\frac{\sinh{kr}}{\sinh{ka}}&r\leq a\\
  P_{\nu}(r)&r>a
  \end{array}\right\}.
\end{equation}\end{itemize}
In Figure (\ref{pic:whitkccsq}) we compare the
wave functions  for both deep and shallow donors, with both $\delta$-function
square-well potential central cell corrections, and
 without any central-cell correction.  Note that in both cases the diverging
tails of the wave function are cut off for $r<a$.

\subsection{Exchange calculations}
Within effective-mass theory, the exchange between two hydrogenic
donors as a function of distance maps to the exchange between two
hydrogen atoms as a function of bond length. This problem has been
treated using different methods since the 1920s.
\subsubsection{The Heitler-London model}
The Heitler-London model \cite{heitlerlondon} simply evaluated the
difference between the expectation values of the Hamiltonian in
two two-electron wave functions, one a singlet ($^1\Sigma_g$) and
one a triplet ($^3\Sigma_u$), both constructed from the 1s ground
states $\phi_a(\vec{r})$ and $\phi_b(\vec{r})$ of the single
atoms.  This approach neglects any contributions from other atomic
states, and hence also contributions to the correlation arising
from the polarization of one atom in the field of the other. The
energy difference can be written as
\begin{equation}\label{eq:heitlerlondon}
J\equiv E_t-E_s={2\over 1-\tS^4}[2j\tS^2+j'\tS^2-2k\tS-k'],
\end{equation}
\begin{eqnarray}
\tS&\equiv&\int \phi_a(1)\phi_b(1)\,\d[1];\\
j&=&\int \phi_a(1)[-{1\over r_{1b}}]\phi_a(1)\,\d[1];\label{eq:defineJ}\\
k&=&\int \phi_a(1)[-{1\over r_{1b}}]\phi_b(1)\,\d[1];\label{eq:defineK}\\
j'&\equiv&\int \phi_a^2(1)\phi_b^2(2){1\over r_{12}}\,\d[1]\d[2];\\
k'&\equiv&\int \phi_a(1)\phi_b(1)\phi_a(2)\phi_b(2){1\over
r_{12}}\,\d[1]\d[2].
\end{eqnarray}
All these quantities are positive; the one-electron contributions
involving $j$ and $k$ yield a net antiferromagnetic coupling
(positive $J$ in our sign convention), while the two-electron
terms $j'$ and $k'$ are net ferromagnetic. However, the neglect of
correlation results in an unphysical logarithmic divergence in
$k'$ at large internuclear separations $R$, causing the overall
$J$ to become negative (ferromagnetic), in violation of the
theorem proved by Heisenberg \cite{heisenberg} showing that the
spatial ground-state of a two-electron system is always even under
exchange of the particle positions, and hence is a spin singlet.

We note that, in order to write $J$ in the form of
equation~(\ref{eq:heitlerlondon}), it is essential that $\phi_a$
and $\phi_b$ be exact eigenstates of the single-atom problem with
energy $E_0=-1/2$. This allows one to replace the off-diagonal
matrix elements of the kinetic energy operator by
\begin{widetext}
 \begin{equation}\label{eq:koillerapprox}
 \int\phi_a(1)[-\frac{1}{2}\nabla_1^2]\phi_b(1)\,\d[1]=\int \phi_a(1)[E_0+{1\over r_{1b}}]\phi_b(1)\,\d[1].
 \end{equation}
\end{widetext}
We refer to this as the `Koiller method', since it is used in \cite{koiller01,koiller02}.

If (\ref{eq:koillerapprox}) is not obeyed  (i.e. if $\phi_a$ and
$\phi_b$ are not exact eigenfunctions of the single-donor
problem), an alternative expression must be used:
\begin{widetext}
\begin{eqnarray}\label{eq:basicformulahl}
\nonumber\bra{\psi_{\mp}}\hat{H}\ket{\psi_{\mp}}&=&\frac{1}{1\mp
S^2}[2\bra{\phi_a(1)}(-\frac{1}{2}\nabla^2)\ket{\phi_a(1)}+2\bra{\phi_a(1)}(-\frac{1}{r_{1A}})\ket{\phi_a(1)}
\\\nonumber&&+2\bra{\phi_a(1)}(-\frac{1}{r_{1B}})\ket{\phi_a(1)}\mp 2S(\bra{\phi_a(1)}(-\frac{1}{2}\nabla^2)\ket{\phi_b(1)}
+2\bra{\phi_a(1)}(-\frac{1}{r_{1A}})\ket{\phi_b(1)})
\\&&+\int
d[1]d[2]\frac{\phi_a^2(1)\phi_b^2(2)}{r_{12}}\mp\int
d[1]d[2]\frac{\phi_a(1)\phi_b(1)\phi_a(2)\phi_b(2)}{r_{12}}]+\frac{1}{R}
\\J&=&\bra{\psi_{-}}\hat{H}\ket{\psi_{-}}-\bra{\psi_{+}}H\ket{\psi_{+}}.
\end{eqnarray}
\end{widetext}
We refer to this as the `exact Heitler-London' method below.

In either case, when using a central-cell correction of the type
discussed in \S\ref{sec:centralcell}, it is important to include
the central-cell terms in the appropriate matrix elements of the
single-particle potential, i.e. in the calculation of the
quantities $j$ and $k$ via equations (\ref{eq:defineJ}) and
(\ref{eq:defineK}), and in the one-electron terms in
(\ref{eq:basicformulahl}).
\subsubsection{Other trial wavefunctions}
Other trial wavefunctions have been proposed for the hydrogenic
case in an attempt to give some account of electron-electron
correlation and remove the unphysical sign change in $J$. Kolos
\textit{eq al.} \cite{kolos} used trial wave functions in
elliptic-coordinates similar to those proposed by \cite{james} to
perform a variational calculation of singlet and triplet energies,
obtaining the exchange splitting from the energy difference.  The
numerical results are compared with the Heitler-London values in
Figure~(\ref{pic:hlhfk}).
\subsubsection{The Herring-Flicker asymptotic form}
A different approach was pursued by Herring \cite{herring62} and
by Gor'kov and Pitaevskii \cite{gorkov63}, who pointed out that
the exchange could be written in terms of a hyper-surface integral
as
\begin{equation}
J={1\over 2}\int_{S}\textbf{dS}[(P\Phi_1)\nabla\Phi_1-\Phi_1\nabla(P\Phi_1)]+\mathrm{O}(\e^{-4R}),
\end{equation}
where
\begin{equation}
\Phi_1=\frac{1}{\sqrt{2}}(\phi_g+\phi_u)
\end{equation}
is a spatial wavefunction that is not properly antisymmetrized and
has electron 1 localized on atom $a$ and electron 2 on atom $b$,
while (assuming the nuclear separation is in the $z$ direction)
the hypersurface $S$ is defined by the condition $z_1=z_2$.

Herring and Flicker \cite{herringflicker} showed that this formula could
be evaluated exactly in the limit $R\rightarrow\infty$, giving an
asymptotic form for the exchange of
\begin{eqnarray}\label{eq:hfex}
J_{\mathrm{H-F}}&=&E_\mathrm{triplet}-E_\mathrm{singlet}\nonumber
\\&=&1.642R^{5/2}e^{-2R}+\textsl{O}(R^2e^{-2R}).
\end{eqnarray}
No unphysical sign change occurs in the Herring-Flicker asymptotic form.

\begin{figure}[htb]
\begin{tabular}{c}
\includegraphics[width=6cm,height=4cm]{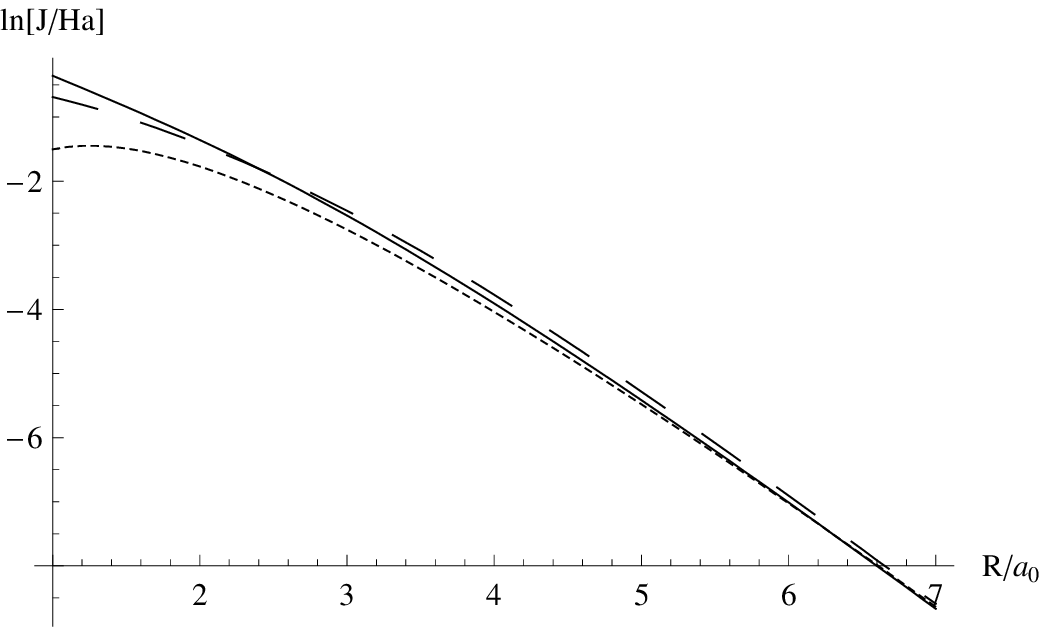}\\
(a)\\
\includegraphics[width=6cm,height=4cm]{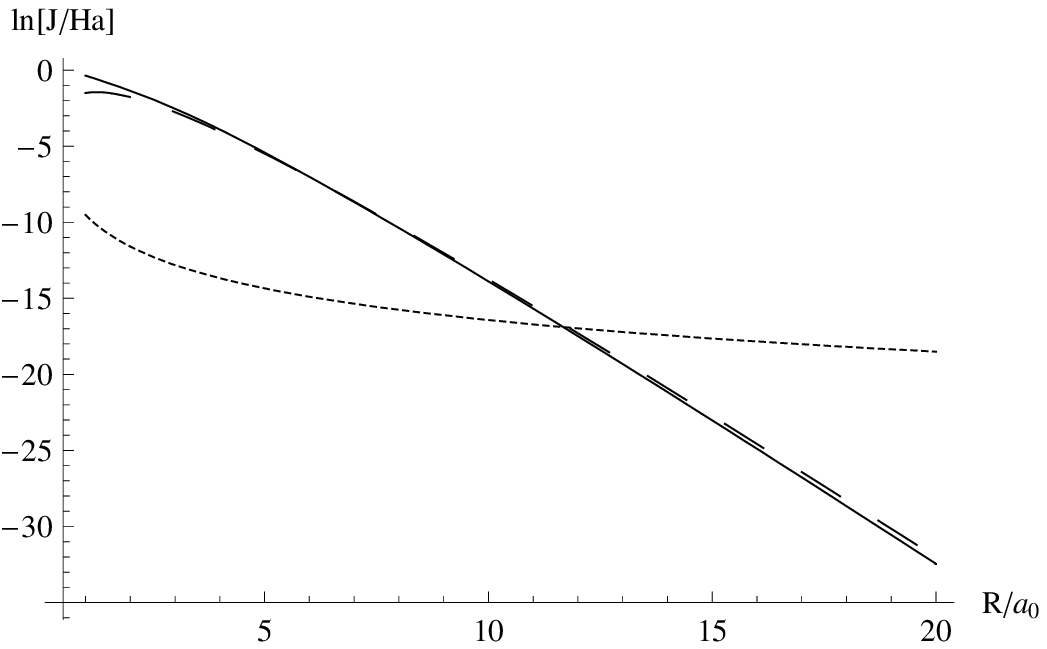}\\
(b)\\
\end{tabular}
\caption{The single-envelope exchange splitting $J$ for hydrogenic
donors calculated by three different methods as a function of
inter-nuclei distance $R$ and shown on a logarithmic scale. (a)
Heitler-London exchange (solid curve), Herring-Flicker asymptotic
(dotted curve) and Kolos' numerical calculation (dashed curve) for
$R\le 7a_0$; (b) Heitler-London exchange (solid curve),
Herring-Flicker asymptotic (dotted curve) and magnitude of
magnetic dipolar interaction for $R\le 20 a_0$. Notice that
Heitler-London model agrees closely with the Herring-Flicker
asymptotic form, and that from around $R=12a_0$ the dipole-dipole
interaction is larger than the exchange
interaction.}\label{pic:hlhfk}
\end{figure}

We compare the Heitler-London, Kolos and Herring-Flicker results
for values of $R$ up to $7a_0$ (using a logarithmic scale) in
Figure~\ref{pic:hlhfk}(a); we also give a comparison for larger
values of $R$ (up to $20a_0$) in Figure~\ref{pic:hlhfk}(b). Note
that all three calculations agree quite well up to $R=7a_0$ (the
maximum radius for which Kolos' numerical results were
calculated), and that the Heitler-London and Herring-Flicker forms
are also close for $R\le 20a_0$. We therefore expect that the
Heitler-London approximation is a reasonable one in this distance
range (the unphysical sign change to a ferromagnetic interaction
does not occur until $R=49.5\,a_0$).  For comparison,
Figure~\ref{pic:hlhfk}(b) also shows the variation of the quantity
$\mu_0 g^2 \mu_B^2/4\pi r^3$, which determines the magnitude of
the magnetic dipolar interaction; we see that the dipolar
interaction starts to dominate over exchange for hydrogenic
defects when $R\ge 12 a_0^*$.

\subsubsection{Inter-valley effects}
The calculations presented so far give the exchange for molecular
hydrogen. When we consider impurity states in a semiconductor we
must remember that the full wavefunction is of the form
(\ref{eq:effectivemasssum}).  Assuming that the intervalley
couplings are mainly determined by single-impurity physics (so
equation~(\ref{eqn:a0toa5}) still holds), the Heitler-London
exchange formula (\ref{eq:heitlerlondon}) should be replaced
\cite{koiller02} by
\begin{eqnarray}\label{eq:kex}
J(\vec{R})&=&\sum_{\nu,\mu}[\sum_{\vec{K},\vec{K}'}\abs{c^{\nu}_{\vec{K}}}^2\abs{c^{\mu}_{\vec{K}'}}
^2e^{i(\vec{K}-\vec{K}') \cdot\vec{R}}]
\\&&\nonumber\times\abs{\alpha_{\nu}}^2\abs{\alpha_{\mu}}^2J_{\nu\mu}(\vec{R})
\cos(\vec{k}_{\nu}-\vec{k}_{\mu})\cdot\vec{R},
\end{eqnarray}
where the pair of donors are at $\vec{R}_A=0, \vec{R}_B=\vec{R}$
and $R>>a_0^*$(effective Bohr radius).  The second sum (in square
brackets) in equation (\ref{eq:kex}) refers to the
reciprocal-lattice expansion of the periodic Bloch functions,
$u_{\nu}(\vec{r})=\sum_{\vec{K}}c^{\nu}_{\vec{K}}e^{i\vec{K}\cdot\vec{r}}$,
and $\vec{k}_{\nu}, \vec{k}_{\mu}$ are band minima points. The
full expression for $J_{\nu\mu}$ is given  in the Appendix of
\cite{koiller02}; in the isotropic effective-mass approximation
where the envelope functions $F_n(\vec{r})$ are the same for each
minimum, and assuming that rapidly oscillating terms in the
integrals (proportional to $e^{i(\vec{k}^{(i)}-\vec{k}^{(j)})\cdot
\vec{r}}$, where $r$ is one of the integrated variables) are
negligible, $J_{\mu\nu}$ can be replaced by the exchange $J_w$
computed using the radial functions derived from the Whittaker
functions. The final expression for the exchange is then
\begin{eqnarray}\label{eqn:wex}
J(R)&=&\sum_{\nu,\mu}[\sum_{\vec{K},\vec{K}'}\abs{c^{\nu}_{\vec{K}}}^2\abs{c^{\mu}_{\vec{K}'}}
^2e^{i(\vec{K}-\vec{K}') \cdot\vec{R}}]
\\&&\nonumber\times\abs{\alpha_{\nu}}^2\abs{\alpha_{\mu}}^2J_w(R)
\cos(\vec{k}_{\nu}-\vec{k}_{\mu})\cdot\vec{R}.
\end{eqnarray}
\subsection{Fitting Whittaker function with $1s$ Gaussian}
Once modified by our model central cell correction, Whittaker
functions  are valid solutions for the single-impurity problem for
the whole range of $r$. To evaluate the integrals appearing in
$J_w(R)$, we expand these solutions as a sum of Gaussians and use
the analytical formulae in \cite{boys}.  We write
\begin{eqnarray}\label{eq:fitwhitk}
\frac{P_{\nu,0}(r)}{r}\simeq R^{G}_{\nu}(r)=\sum_{n}A_{n}G_{n}(r,B_{n})\\
\nonumber G_{n}(r,B_{n})=e^{-B_{n}r^2}.
\end{eqnarray}
In all the calculations presented here, we use \textsl{ten} $1s$
Gaussian-type orbitals to fit each Whittaker function.

\begin{figure}[htb]
\begin{center}
\begin{tabular}{c}
\includegraphics[width=6cm,height=4cm]{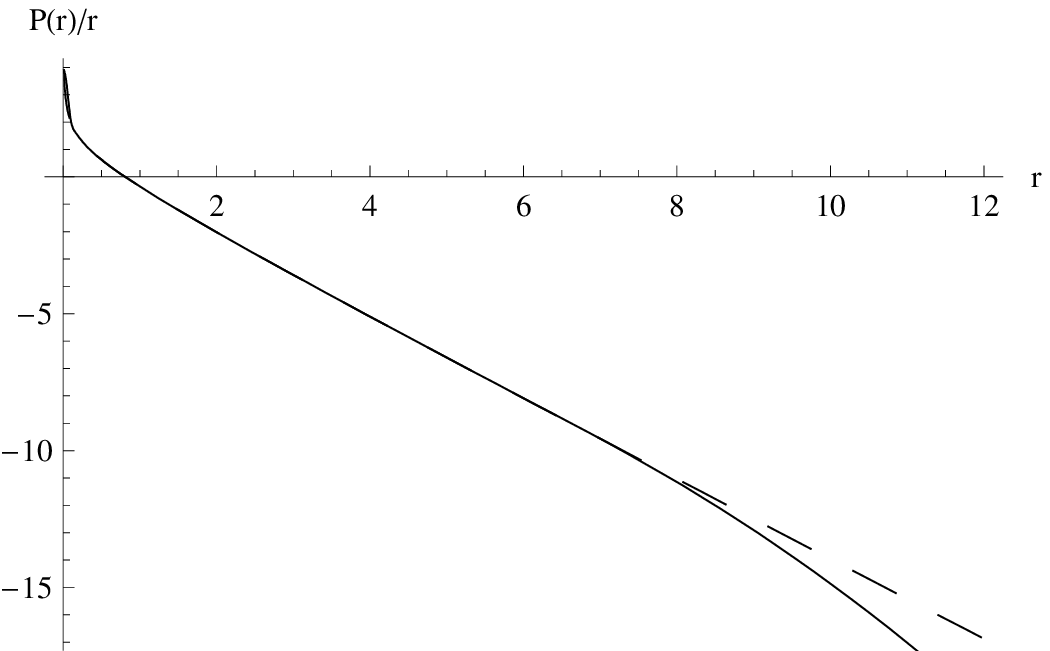}\\
(a)\\
\includegraphics[width=6cm,height=4cm]{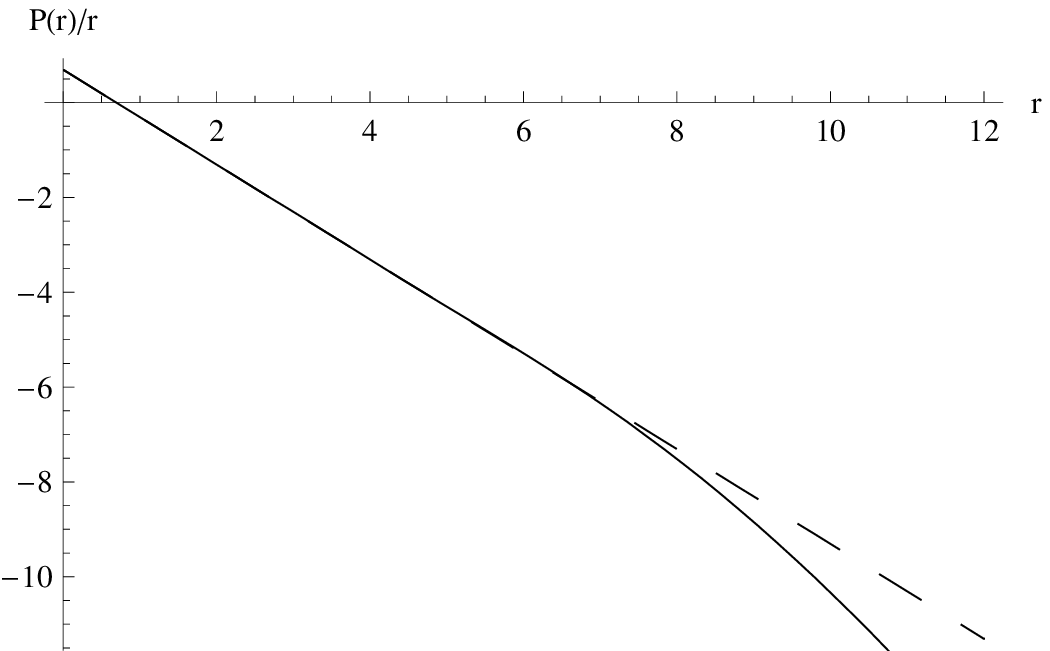}\\
(b)\\
\includegraphics[width=6cm,height=4cm]{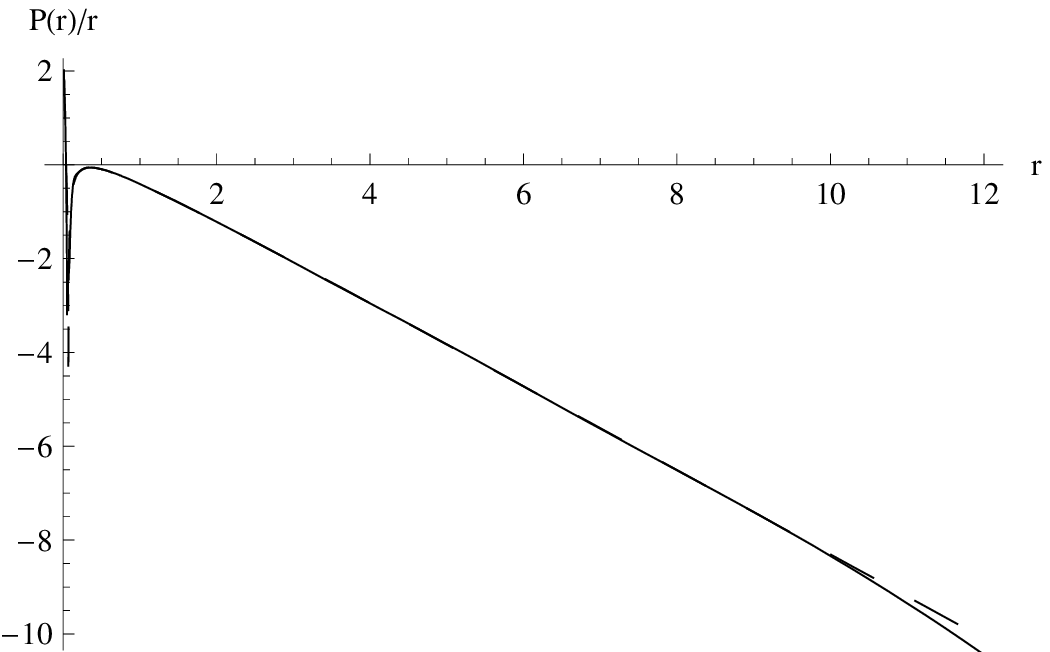}\\
(c)\\
\end{tabular}
\caption{Gaussian fits of Whittaker functions for (a) typical deep
donor ($\nu=0.7$), (b) typical hydrogenic shallow donor electron
($\nu=1.0$) and (c) `super-shallow' donor ($\nu=1.1$). In each
case the solid curve is the Gaussian fit and the dashed curve is
the Whittaker function $P(r)/r$.}\label{pic:fit2}
\end{center}
\end{figure}

In Figure~(\ref{pic:fit2}) we show comparisons of such fits with
the original wave functions. We see that we obtain a very good fit
over the physically interesting distance range $R\le 12 a_0^*$.

\section{Results}

\subsection{Choice of parameters}
%Give experimental binding energies for important species
We select parameters appropriate for donors in silicon, as
described previously: the appropriate effective mass is
$m^*=(m_\perp^2*m_\parallel)^\frac{1}{3}=0.33m_e$ and the relative
permittivity $\epsilon_r=11.7$, leading to an effective Bohr
radius $a_0^*=0.053\mathrm{nm}{{\epsilon_r m^*}\over m_e}=1.94
\mathrm{nm}$ and an effective Hartree
$\mathrm{Ha}^*=27.2\,\mathrm{eV}(a_0/a_0^*)=62 \ \mathrm{meV}$.

We take as a typical deep donor the case of Bi:Si, for which the
binding energy is about $71 \mathrm{meV}$ \cite{ramdas81} or
approximately $1.15\ \mathrm{Ha}^*$; according to
formula~(\ref{eq:donorenergy}), we therefore adopt $\nu=0.7$ as a
typical value for a deep donor, as used in
Figure~(\ref{pic:whitkccsq}). For comparison and to show that our
method is also capable of dealing with the case $\nu > 1$, we also
show results for the case $\nu=1.1$, corresponding to a
hypothetical `super-shallow' donor whose binding energy is
\emph{less} than the effective-mass value.  (This is of less
practical interest, as the shallowest known donor, Li:Si, has
$\nu$ almost exactly equal to 1.)  In all case where a model
central-cell correction (of either type) is used, the radius is
set to $a=0.01 a_0^*$.

The results presented here cover the distance range from $a_0^*$
to $12 a_0^*$. This range includes the lengthscales of greatest
interest in practice, both for studies of the metal-insulator
transition and for applications in quantum information processing,
since it runs from the typical nearest-neighbour separation at the
highest attainable densities below the metal-insulator transition
(at a donor density $n_D=4\times 10^{18}\,\mathrm{cm}^{-3}$, the
mean inter-donor spacing is $3.5\,\mathrm{nm}$ or $1.8\,
a_0^*$ and approximately 90\% of nearest-neighbour separations are
greater than $a_0^*$) up to separations where the dipole-dipole
interaction starts to dominate over exchange (see Figure~\ref{pic:hlhfk}).

\subsection{Benchmark: hydrogenic donors, $\nu=1.0$}
\begin{figure}[htb]
\begin{center}
\includegraphics[width=9cm,height=8cm]{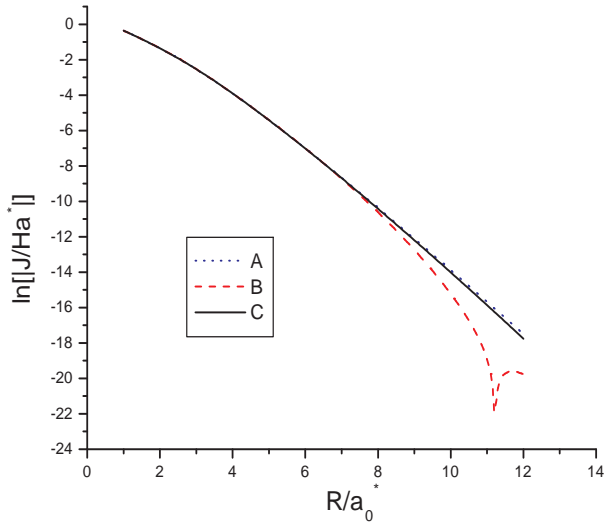}
\caption{(Colour online.)  The exchange coupling between
hydrogenic donors as a function of inter-nuclear distance $R$.
Blue dotted curve (\textbf{A}): exchange calculated directly from
Hydrogenic $1s$ Coulomb wave function; red dashed curve
(\textbf{B}): exchange from Koiller's method based on
$R^{G}_{1.0}$; black solid curve (\textbf{C}): exact
Heitler-London method from $R^{G}_{1.0}$. }\label{pic:exc1}
\end{center}
\end{figure}

First, we perform a benchmark calculation in which we compute the
Heitler-London exchange interaction between two pure hydrogenic
donors (i.e. $\nu=1$) by two different methods: first using the
standard evaluation of the integrals from the exact Coulomb
wavefunctions, and second using our Gaussian fit.  The results are
shown in Figure~(\ref{pic:exc1}); the excellent agreement between
the dotted line \textbf{A} (exact Coulomb wavefunctions) and the solid line \textbf{C}
(Gaussian fit) shows that the errors introduced by the Gaussian
fit are negligible.  There is however some deviation at large
distances ($R\ge 8 a_0^*$) between the solid curve \textbf{C} (exact
Heitler-London result) and the dashed curve \textbf{B} (Koiller method); this
arises because our Gaussian fit is not an exact eigenfunction of
the one-centre problem and culminates in an unphysical cusp in curve \textbf{B} (corresponding to a sign-change in the exchange) at $R\approx 11a_0$.

\subsection{Deep donors, $\nu= 0.7$}
\begin{figure}[htb]
\begin{center}
\includegraphics[width=9cm,height=8cm]{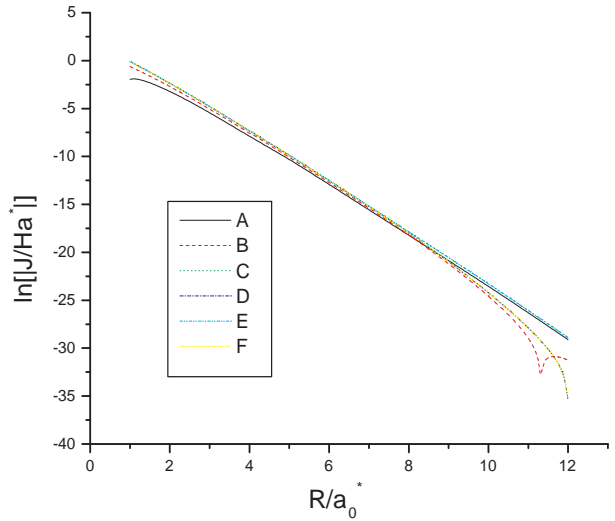}
\caption{(Colour online.) The exchange coupling between two deep
donor electrons ($\nu=0.7$) as a function of inter-donor distance
$R$. Black solid curve (\textbf{A}): exact Heitler-London method
without central cell correction; red dashed curve (\textbf{B}):
Koiller's method without central cell correction; green dotted
curve (\textbf{C}): exact Heitler-London method with
$\delta$-potential central cell correction; blue dashed-dotted
curve (\textbf{D}): Koiller's method with $\delta$-potential
central cell correction; cyan dashed-dotted-dotted curve
(\textbf{E}): exact Heitler-London method with square-well central
cell correction; yellow short-dashed curve (\textbf{F}): Koiller's
method with square-well central cell correction. (\textbf{D} and
\textbf{F} are indistinguishable at long range).  The cusp in curve \textbf{B}, corresponding to a sign change at $R\approx 11a_0^*$ is unphysical (see
text).}\label{pic:deep07}
\end{center}
\end{figure}

Now we know that the errors arising from our Gaussian fits are
likely to be small in the region of interest, we can move on to
calculate our main result: the exchange coupling between two deep
donors.  The results for $J_w$, calculated from a single envelope
function, are shown in Figure~(\ref{pic:deep07}). Notice that the
six curves shown have very similar behaviours over most of the
distance range. The three curves calculated using the `exact
Heitler-London method' and the different model central-cell
corrections (\textbf{A, C, E}) are very close to one another, but
in this case there are some significant deviations among the
curves calculated using  `Koiller's method'(\textbf{B, D, F}),
both from the exact Heitler-London results and among one another.
These deviations arise because the left side and right sides in
equation~(\ref{eq:koillerapprox}) are not exactly equal: in the
cases of curves \textbf{D} and \textbf{F}, where central-cell
corrections are included, this discrepancy arises from errors in
the Gaussian fit, whereas in the case of curve \textbf{B} (no
central-cell correction) there is an additional error because the
Whittaker function is not a true eigenfunction of the potential.
This introduces a larger error and leads to an (unphysical)
cusp corresponding to a sign-change in the predicted exchange at $R\approx 11a_0^*$

\begin{figure}[htb]
\begin{center}\begin{tabular}{cc}
\includegraphics[width=4cm,height=4cm]{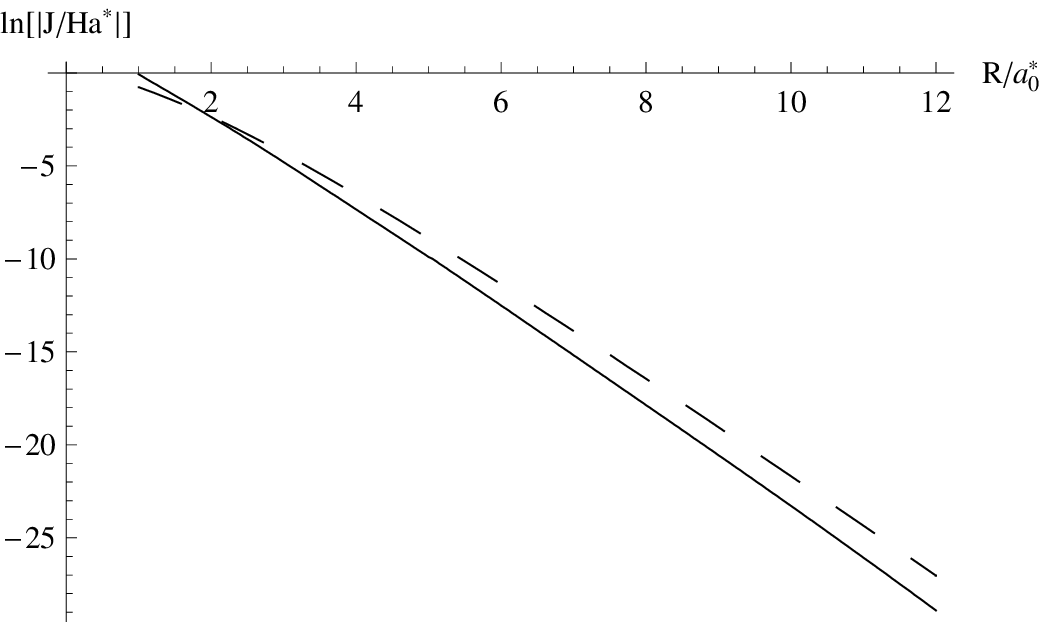}&\includegraphics[width=4cm,height=4cm]{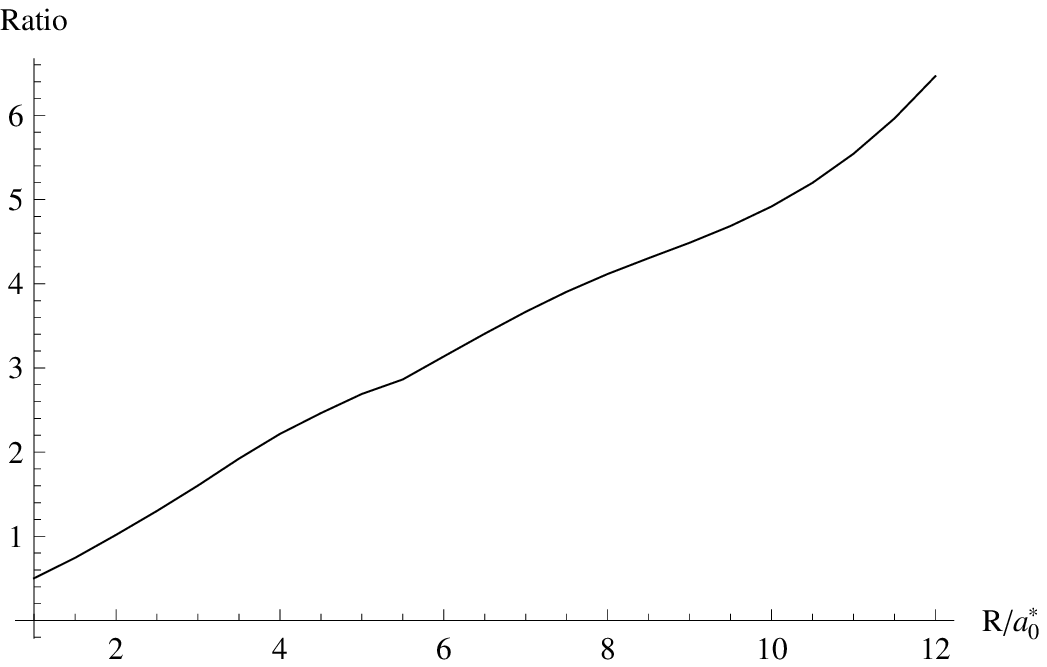}\\
(a)&(b)
\end{tabular}
\caption{Comparison of exact Heitler-London exchange for deep
donors ($\nu=0.7$) with scaled hydrogenic results.  (a) Exact
Heitler-London exchange $J_\mathrm{exact}$ based on $R^{G}_{0.7}$
with a square-well potential central-cell correction (solid line)
and scaled exchange splitting using the hydrogenic Heitler-London
formula $J_\mathrm{scaled}=J_\mathrm{H-L,\, hydrogenic}(R/0.7)$
(dashed line). (b) The ratio
$\frac{J_\mathrm{scaled}}{J_{\mathrm{exact}}}$ as a function of
inter-donor distance.}\label{pic:scaledex07}
\end{center}
\end{figure}

Comparing the computed exchange with the results for hydrogenic
states (Figure~\ref{pic:exc1}), we see that the dominant effect is
the change in exponential decay constant reflecting the change in
the exponent of the radial wavefunctions at large distances.  It
is therefore natural to ask whether a good approximation to
exchange in deep donors can be obtained simply by re-scaling the
results from the Heitler-London approach for a hydrogenic donor,
making the replacement $R\rightarrow R/\nu$. This comparison is
made in Figure~\ref{pic:scaledex07}(a); we see that the scaled
interactions have qualitatively the right behaviour and the
correct order of magnitude, but do not match the details of our
calculation well. This impression is confirmed by looking at the
ratio of the scaled to exact results shown in
Figure~\ref{pic:scaledex07}(b): the error introduced by using the
scaled approximation is nearly one order of magnitude over the
distance range shown here, and increases still further at larger
inter-donor distances.

\subsection{`Super-shallow' donors, $\nu=1.1$}
\begin{figure}[htbp]
\begin{center}
\includegraphics[width=9cm,height=8cm]{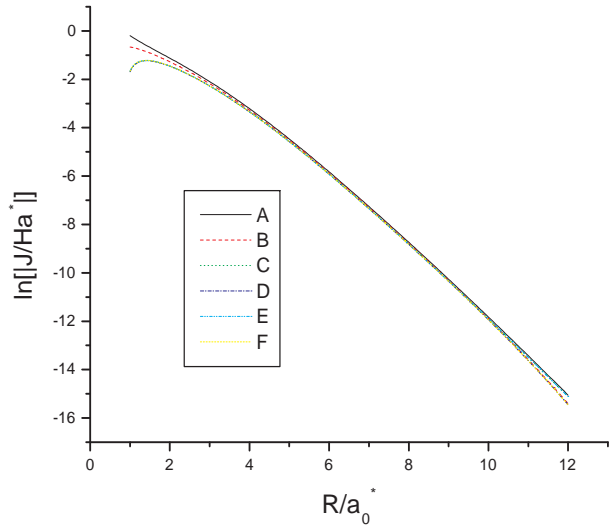}
\caption{(Colour online.)  The exchange coupling between two super-shallow donor
electrons ($\nu=1.1$) as a function of inter-donor distance $R$. Black solid
curve (\textbf{A}): exact Heitler-London method without central
cell correction; red dashed curve (\textbf{B}): Koiller's method
without central cell correction; green dotted curve (\textbf{C}):
exact Heitler-London method based on with $\delta$-potential
central cell correction; blue dashed-dotted curve (\textbf{D}):
Koiller's method with $\delta$-potential central cell correction;
cyan dashed-dotted-dotted curve (\textbf{E}): exact Heitler-London
method with square-well central cell correction; yellow
short-dashed curve (\textbf{F}): Koiller's method with square-well
central cell correction. Notice that these six curves have very
similar behaviour over the whole range except at short distance
where there is a dependence on the central cell
corrections.}\label{pic:exc11}
\end{center}
\end{figure}

%In this step we change the basic functions to $P_{1.1,0}(r)$.
%\begin{eqnarray}
%\psi_{A}=\frac{1}{\sqrt{4\pi}}P_{1.1,0}(\abs{\vec{r}-\vec{R_{A}}})/\abs{\vec{r}-\vec{R_{A}}};
%\nonumber
%\\\psi_{B}=\frac{1}{\sqrt{4\pi}}P_{1.1,0}(\abs{\vec{r}-\vec{R_{B}}})/\abs{\vec{r}-\vec{R_{B}}};
%\nonumber\\\abs{\vec{R_{A}}-\vec{R_{B}}}=R.
%\end{eqnarray}
We also show the single-envelope exchange couplings $J_w(R)$ in the case of two
`super-shallow' donors. The results are shown in
Figure~(\ref{pic:exc11}). Once again the results are independent
of the type of central-cell correction used; furthermore (in the range $R\le
12a_0^*$) they are
now largely independent of whether the exact Heitler-London
approach or the `Koiller method' is used .

\begin{figure}[htb]
\begin{center}\begin{tabular}{cc}
\includegraphics[width=4cm,height=4cm]{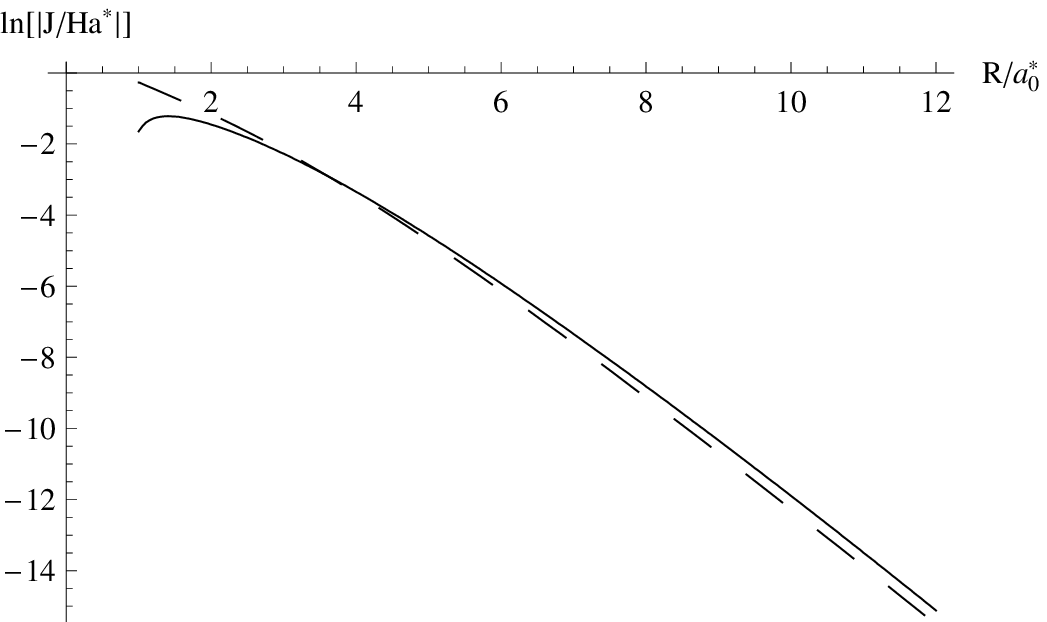}&\includegraphics[width=4cm,height=4cm]{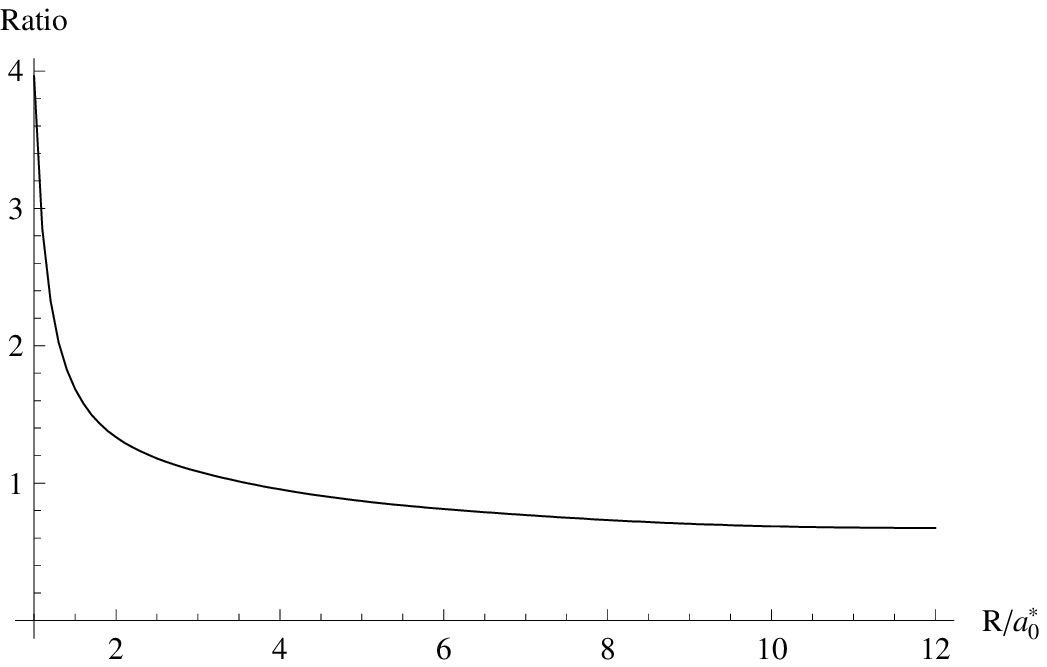}\\
(a)&(b)
\end{tabular}
\caption{Comparison of exact Heitler-London exchange for
super-shallow donors ($\nu=1.1$) with scaled hydrogenic results.
(a) Exact Heitler-London exchange $J_\mathrm{exact}$ with a square-well potential
central-cell correction (solid curve) and scaled exchange splitting
using the hydrogenic Heitler-London formula $J_\mathrm{scaled}=J_\mathrm{H-L,\, hydrogenic}(R/1.1)$ (dashed curve). (b)
The ratio $\frac{J_\mathrm{scaled}}{J_{\mathrm{exact}}}$ as a function of inter-donor
distance.}\label{pic:scaledex11}
\end{center}
\end{figure}

Comparing the results to the hydrogenic case
(Figure~\ref{pic:exc1}), the dominant difference is once again the
change in the exponent.  As previously, we compare with the scaled
hydrogenic Heitler-London result. Figure~\ref{pic:scaledex11}(a)
shows that the matching is much closer than for the deep donors;
the relative error is shown in Figure~\ref{pic:scaledex11}(b). In
contrast to the deep-donor case, the scaled hydrogenic
wavefunction overestimates exchange interaction at short range,
but underestimates it at long range.  It is perhaps not surprising
that the error is smaller in this case, since $\nu$ is closer to
the hydrogenic value $\nu=1$ and so the scaling has a smaller
effect.

\subsection{Inter-valley effects}
\begin{figure}[htb]
\begin{center}
\includegraphics[width=9cm,height=8cm]{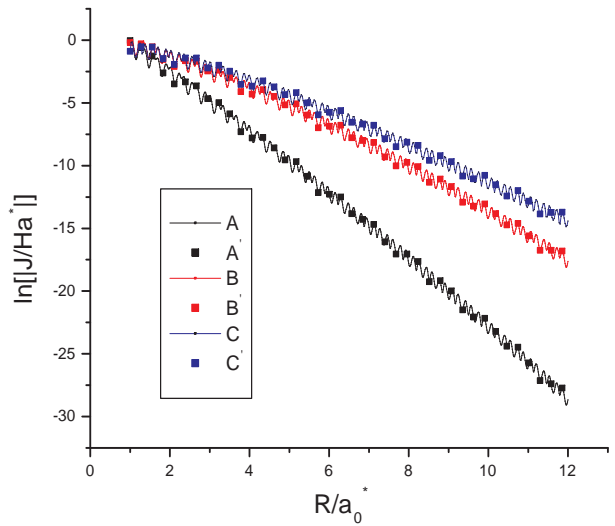}
\end{center}
\caption{(Colour online.)  Donor-pair exchange including
inter-valley effects for pair separations along the $\{100\}$
direction as a function of donor distance R. The exchange at
integer multiples of the lattice constant is represented by square
points. A square-well central-cell correction and the exact
Heitler-London approach were used for all the calculations.  Lower
(black) group (\textbf{A, A'}): two deep donors with $\nu=0.7$;
middle (red) group (\textbf{B, B'}): two typical shallow donors
with $\nu=1.0$; the upper (blue) group (\textbf{C, C'}): two
super-shallow donors with $\nu=1.1$.}\label{pic:pic3fortalk}
\end{figure}

Having calculated $J_w(R)$, we include the inter-valley effects by
using equation~(\ref{eqn:wex}). We have calculated the exchange
between typical deep donors ($\nu=0.7$), shallow donors
($\nu=1.0$), and super-shallow donors ($\nu=1.1$) in silicon in
the case where the donor pair is along a $\langle 100\rangle$
direction from $1 a_0^*$ to $12 a_0^*$ and both donors are in the
ground state ($A_1$ symmetry, so equation~(\ref{eqn:a0toa5})
applies). We see the exchange interactions still decay
exponentially over the whole range, but this decay is mixed with
oscillations due to the inter-valley terms as shown in
Figure~(\ref{pic:pic3fortalk}).

The solid points in the figure denote the distances corresponding
to the cubic lattice constant in Si (i.e., the actual separations
of substitutional sites along $[100]$). For all three defect
types, the interference terms produce deviations from site to site
of approximately one order of magnitude, in agreement with the
results for hydrogenic defects \cite{koiller01}.

\subsection{Distributions of exchange couplings}
\begin{figure}[htb]
\begin{center}
\includegraphics[width=8cm,height=8cm]{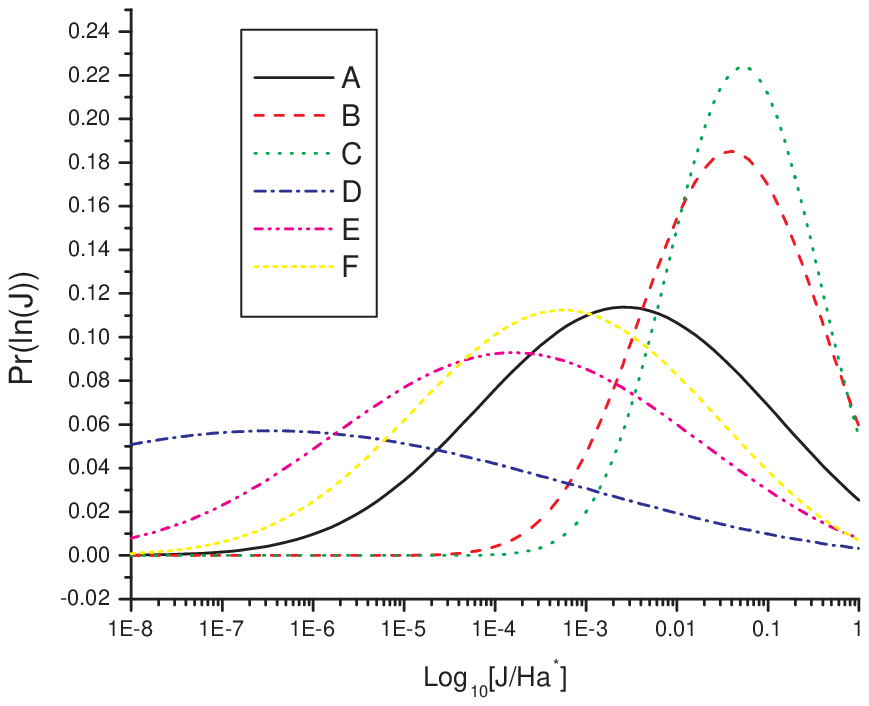}
\end{center}
\caption{(Colour online.) The probability distribution of $\log J$
as a function of $J$ (logarithmic scale) for donors with three
different binding energies at densities $0.00396\,(a_0^*)^{-3} =
5.42\times 10^{17}\,\mathrm{cm}^{-3}$ (curves \textbf{A, B, C})
and $0.0005\,(a_0^*)^{-3} = 6.85\times 10^{16}\,\mathrm{cm}^{-3}$
(curves \textbf{D, E, F}). Solid (black) curve \textbf{A} and
dashed-dotted (blue) curve \textbf{D}: deep donor $\nu=0.7$;
dashed (red) curve \textbf{B} and dashed-dotted-dotted (magenta)
curve \textbf{E}: shallow donor $\nu=1.0$; dotted (green) curve
\textbf{C} and short-dashed (yellow) curve \textbf{F}:
super-shallow donor $\nu=1.1$.}\label{pic:pj071011}
\end{figure}

In the interpretation of experiments on the ensemble of
interacting spins, it is frequently the probability distribution
of the nearest-neighbour exchange couplings which is the quantity
of most relevance \cite{andres,bhatt}.  In order to show how this
depends on the donor type, we have used a continuous distribution
\cite{chand} of donor positions at two different densities, chosen
so that the mean nearest-neighbour separations are respectively
$3.5 a_0^*\approx 6.8 \mathrm{nm}$ and $7.0 a_0^*\approx
13.5\,\mathrm{nm}$, and plotted the distributions of $\log J$ for
the three different types of defects in
Figure~(\ref{pic:pj071011}).  (The lower limit plotted corresponds
approximately to the values of $J$ where dipolar interactions
begin to dominate over exchange.)  All the distributions are
extremely broad (as would be expected from the exponential scaling
of the exchange with separation) but the strong dependence on the
type of donor is evident.

This distribution is important for the choice of donor
concentration in samples for applications in quantum information
processing; for example, in the scheme proposed in \cite{sfg} the
density should ideally be sufficiently low that typical exchange
interactions between neighbours produce a small evolution of the
system on the timescales of gate operation.  For deep donors with
$\nu=0.7$, typical nearest-neighbour interactions at the lower
density are $10^{-4}\,\mathrm{Ha}^*$ or smaller, constraining gate
operation times to be at most $0.1\,\mathrm{ns}$.

\section{Conclusion}
We have shown that a combination of quantum-defect theory (where
Whittaker functions are used to describe simply the long-range
part of donor wavefunctions) with model central-cell corrections
can be used to describe the electronic structure of donors in
semiconductors having binding energies significantly different
from the ideal hydrogenic effective-mass value, including in
particular the important case of deep donors. From these
wavefunctions we have shown that it is possible to calculate the
exchange interactions between donors, both in the single
envelope-function approximation and by including the interference
effects between contributions from different conduction-band
minima.  These interference effects typically cause the exchange
to fluctuate by approximately one order of magnitude between the
separations of successive substitutional sites, in agreement with
previous calculations for shallow donors.

From the comparison between the exchange calculations performed
with and without different central cell corrections, we can see
that as expected the correct long-range behaviour of the
wavefunction is more important than the form near the nucleus,
except when the inter-donor distance becomes very small---however,
the Heitler-London like approach we use for exchange is not
expected to be accurate at short range.  At long range, the
central-cell corrections have only small effects on the exchange,
although these may be enhanced by the interference of inter-valley
terms.  However, the inclusion of central-cell corrections is
important to obtain consistent results for the exchange between
the exact Heitler-London approach
(equation~(\ref{eq:basicformulahl})) and the frequently-used
simplification given by equation~(\ref{eq:heitlerlondon}).

Our calculations rely on a Gaussian fit to the true form of the
radial function, which is least accurate very near the nucleus and
at large distances.  At large donor separations we expect the
error due to corrections near the nucleus to scale as
$\mathrm{O}(a^3)$, where $a$ is the starting radius for the
fitting. In our calculations $a=0.01 a_0^*$, much less than the
donor separations $1-12a_0^*$ that we consider, so we expect this
error to be negligible.  The long-range error in the fit is
important only at very large donor separations where the spin-spin
interaction is no longer exchange-dominated.

We have shown that the type of defect has a significant effect on
the distribution of nearest-neighbour exchange couplings
experienced at a given density.  To summarize the magnitude of the
effect, at at separation of $3.5\,a_0^*\approx6.8\,\mathrm{nm}$
the exchange is approximately $0.15\,\mathrm{meV}$ for typical
deep donors, $3\,\mathrm{meV}$ for hydrogenic shallow donors, and
$5\,\mathrm{meV}$ for the so-called `super-shallow' donors. This
emphasizes the usefulness of deep donors for the short-term
storage of quantum information, and confirms them as candidates
for quantum information processing provided that entangling
interactions between them can be switched sufficiently quickly
\cite{kane,sfg}.

We have also shown that simply scaling the hydrogenic exchange
interaction from the Heitler-London formula does not agree
quantitatively with our explicit calculations, though it does give
the correct order of magnitude for the exchange. The error is most
serious for the important case of deep donors.

We believe the methods we present here are quantitatively
reliable, and simple enough to provide a useful tool for
calculating exchange interactions between donor pairs having
arbitrary binding energies.

\begin{acknowledgments}
WW was supported by the Research Councils Basic Technology
programme under grant GR/S23506/01.  We thank Thornton Greenland,
Tony Harker, Andy Kerrdidge, Marshall Stoneham and Dan Wheatley
for helpful discussions.
\end{acknowledgments}

%\newpage

\end{document}